\documentclass[a4paper,11pt]{article}
\pdfoutput=1 

\usepackage{jcappub} 

\usepackage[T1]{fontenc} 

\usepackage{listings}
\usepackage{color}

\makeatletter
\def\BState{\State\hskip-\ALG@thistlm}
\makeatother

\definecolor{dkgreen}{rgb}{0,0.6,0}
\definecolor{gray}{rgb}{0.5,0.5,0.5}
\definecolor{mauve}{rgb}{0.58,0,0.82}

\providecommand{\e}[1]{\ensuremath{\times 10^{#1}}}
\def\mpl{M_\mathrm{Pl}}

\usepackage{xspace}

\newcommand{\dd}{\mathrm{d}} 

\newcommand{\ud}{\mathrm{d}}
\newcommand{\pd}{\partial}

\def\bk{{\mathbf{k}}}

\def\calR{{\cal R}}

\def\calS{{\cal S}}

\def\PIJ{{\mathcal{P}_{\delta \phi}^{IJ}}}
\def\Pten{{\mathcal P_h}}

\def\Ppnad{{\mathcal P_{\delta P, \mathrm{nad}}}}

\def\PR{{\mathcal P_\calR}}
\def\PS{{\mathcal P_\calS}}

\title{\boldmath Oscillatory power spectrum and strongly $k$-dependent $r$ in hybrid inflation}

\author{R. Kabir \note{Corresponding author.}}
\author{and A. Mukherjee}

\affiliation{Department of Physics and Astrophysics, University of Delhi, Delhi-110007, India}

\emailAdd{rakesh.kabir@nic.in}
\emailAdd{am@physics.du.ac.in}

\abstract{For the original hybrid inflation model, we calculate the power spectrum of curvature perturbations generated during the waterfall, taking into account the contribution of entropic modes. We study the regime where the potential is very flat, and in which inflation continues for more than about 60 e-folds.  Results show oscillations in the scalar power spectrum, qualitatively similar to that of a single field axion monodromy potential; however no feature is displayed in the tensor power spectrum. Directions to develop a template for this power spectrum are also discussed. }

\begin{document}
\maketitle
\flushbottom

\section{Introduction}~\label{sec:introduction}

Although the inflationary paradigm is successful for solving various problems of the Standard cosmology, it has established itself mainly because of providing seeds for the CMB anisotropy~\cite{Lyth2009, Bardeen1980}. Various single field models of inflation have been shown to be compliant to observations, such as those of WMAP and Planck~\cite{Hinshaw2013, Collaboration2014}. But from the model building point of view, multifield models naturally arise in fundamental theories of High Energy physics, such as SUSY or SUGRA in which occurrence of more than one field is the norm rather than a construct~\cite{Halyo1996, Dvali1994}. Power spectrum of multifield models is generally relegated to the $\delta N$ formalism which relies only on the background solution~\cite{Huston2013, Frazer2014, Clesse2014}, instead of direct solution of perturbation equations. Only numerical techniques can explore the full prediction for power spectrum of multifield model by evolving the full set of perturbation and background equations numerically~\cite{Price2015}. 

Although sum-separable potentials, such as $N_f$-flation etc., can be handled through the standard numerical techniques~\cite{Huston2013}, however, the situation becomes complicated in the case when there is interaction between the fields, such as hybrid inflation~\cite{Wang1994,Clesse2009}. Interaction between the fields turns the system of differentials equations into stiff and chaotic, \textit{i.e.} sensitive to the initial conditions~\cite{Easther2013}. For example in hybrid model, it is quite difficult to find a successful background trajectory which gives 60 e-folds of inflation, and it is known as initial condition problem in literature~\cite{Clesse2009, Easther2013}. The problem becomes more intricate if one desires to find a power spectrum for a successful background trajectory in the hybrid model. In this paper, we have calculated the power spectrum for a successful background trajectory in the hybrid model by directly evolving perturbation equations.

The paper is organized as follows. In Section~\ref{sec:setup} we review two versions of the potential for hybrid  inflation and provide the relation between them, and its effective one-field dynamics is quickly reviewed. More importantly linear theory for multi-field perturbation and power spectra is introduced.  In Section~\ref{sec:numerics} we describe our numerical results.  In Section~\ref{sec:conclusion} we discuss the implication of our results and identify future lines of enquiry.

\section{Formal set up}\label{sec:setup}

Among many ways to realize hybrid inflation, we focus mainly on the two versions studied in~\cite{Clesse2009}  and~\cite{Easther2013}, and  provide the relations between the parameters of both versions. In the first version~\cite{Easther2013}, the potential in field space $(\psi, \phi)$, where $\phi$ is the inflaton and $\psi$ is the waterfall field, is 
\begin{equation}
	\label{eqn:hybridv}
  V(\phi,\psi) = \Lambda^4 \left[\left(1-\frac{\psi^2}{M^2} \right)^2 + \frac{\phi^2}{\mu^2}
  + \frac{\phi^2 \psi^2}{\nu^4} \right],
\end{equation}

with real parameters $\Lambda$, $M$, $\mu$, and $\nu$.

In the second version\cite{Clesse2009}, the potential is 

\begin{equation} \label{eqn:potenhyb2d}
V(\phi,\psi) = \frac 1 2 m^2 \phi^2 + \frac \lambda 4 \left(\psi^2
- M^2 \right)^2 +\frac{\lambda'}{2} \phi^2 \psi^2,
\end{equation}
with real parameters $\lambda$, $\lambda'$ $M$, and $m$. The relations between the parameters of the potentials are as follows: To go from the first to the second 

\begin{equation}
m= \frac{\sqrt{2} \Lambda ^2}{\mu },\text{$\lambda $}= \frac{2 \Lambda ^4}{\nu ^4},\text{$\lambda' $}= \frac{4 \Lambda
   ^4}{\text{M}^4},
\end{equation}
and to go from the second to the first 

\begin{equation}
\Lambda = \frac{\sqrt[4]{\text{$\lambda' $}} \text{M}}{\sqrt{2}},\mu = \frac{\sqrt{\text{$\lambda' $}} \text{M}^2}{\sqrt{2} m},\nu =
   \frac{\sqrt[4]{\text{$\lambda' $}} \text{M}}{\sqrt[4]{2} \sqrt[4]{\text{$\lambda $}}},
\end{equation}
where $M$ is the same in the both the versions.

 \begin{figure}
     \includegraphics{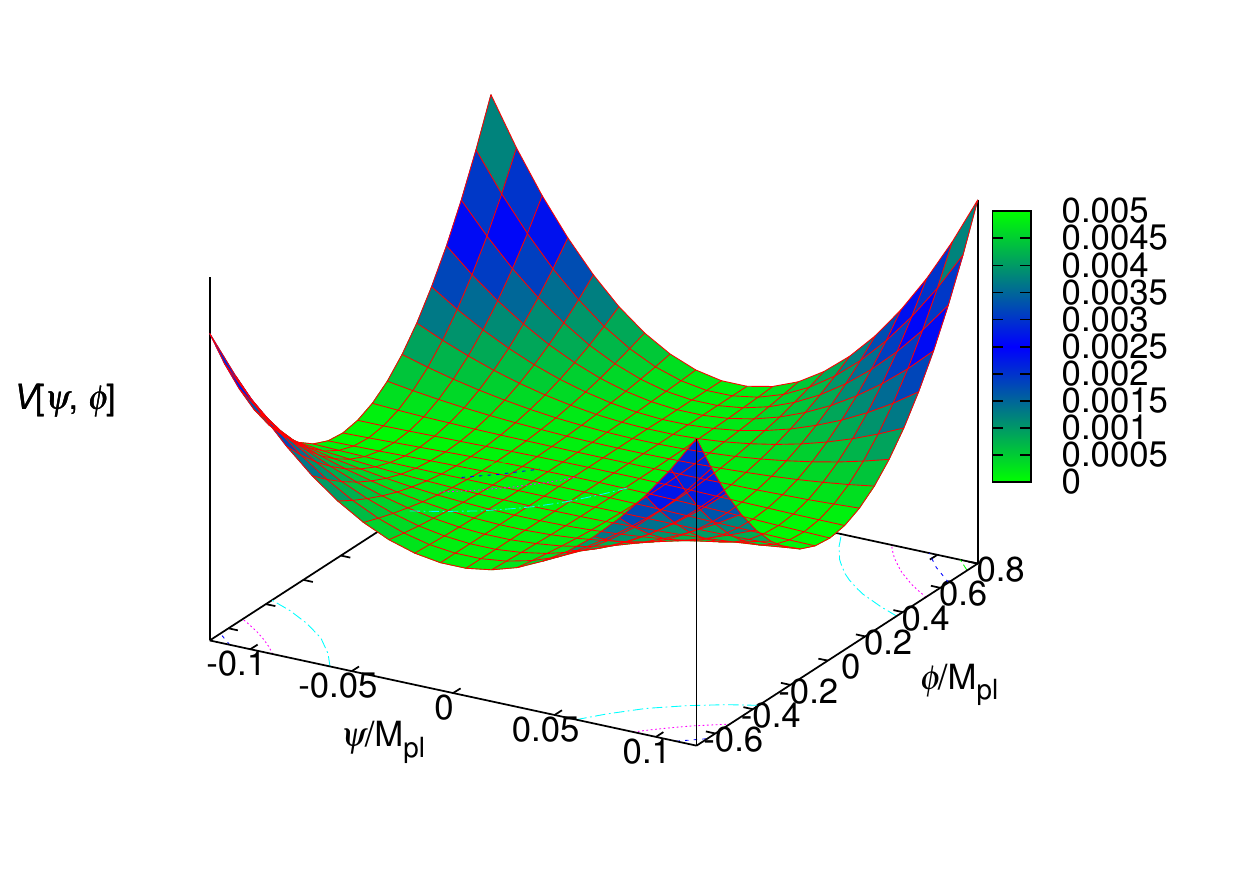}
         \caption{\label{fig:potential} Potential of hybrid inflation with contour in the $(\psi, \phi)$ plane. The  values of the parameters are $\Lambda = 0.1064 \, \mpl$, $M=0.1504  \, \mpl$, $\mu=3190.4  \, \mpl$, and $\nu=0.1265 \, \mpl$  with $\mpl$ being the reduced Planck mass.}
 \end{figure}
 
 For convenience, in what follows, we will consider only the first version of the potential~\eqref{eqn:hybridv}.
  The true minima of the potential are located at $\phi=0$ and  $\psi=\pm M$, while the instability point is given by $\psi=0$, 
 $\phi=\phi_\mathrm{c}$, where  $\phi_\mathrm{c}= \sqrt{2} \, \nu^2/M$ is the critical point at which the effective mass of $\psi$ becomes complex and inflation is assumed to come to an end. Along the inflationary valley, $\psi \approx 0$, the potential reduces to  $\Lambda^4\left[1+(\phi/\mu)^2\right]$ which shows that, in this  regime, inflation cannot end by violation of the slow-roll
 conditions.
 
  \begin{figure}[!t]
  \includegraphics{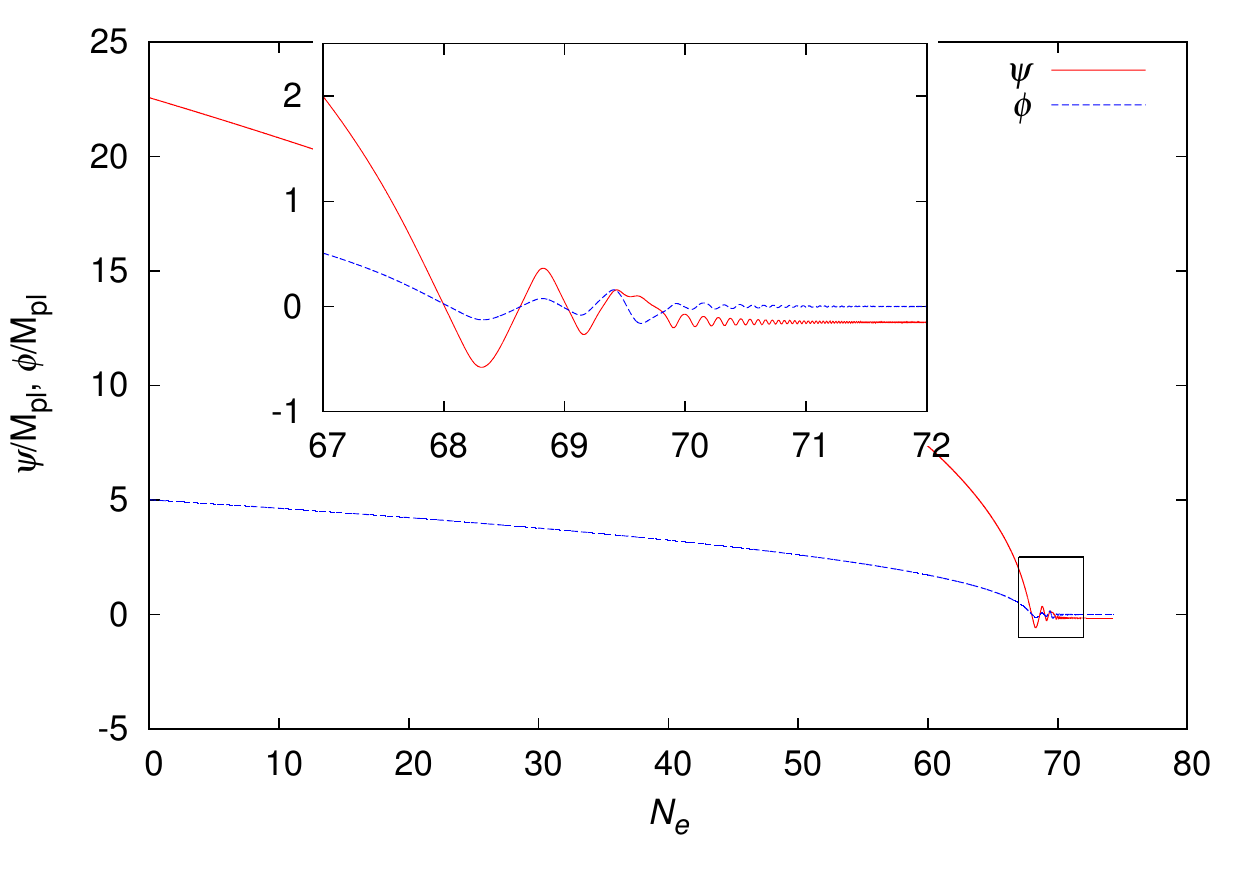} 
  \includegraphics{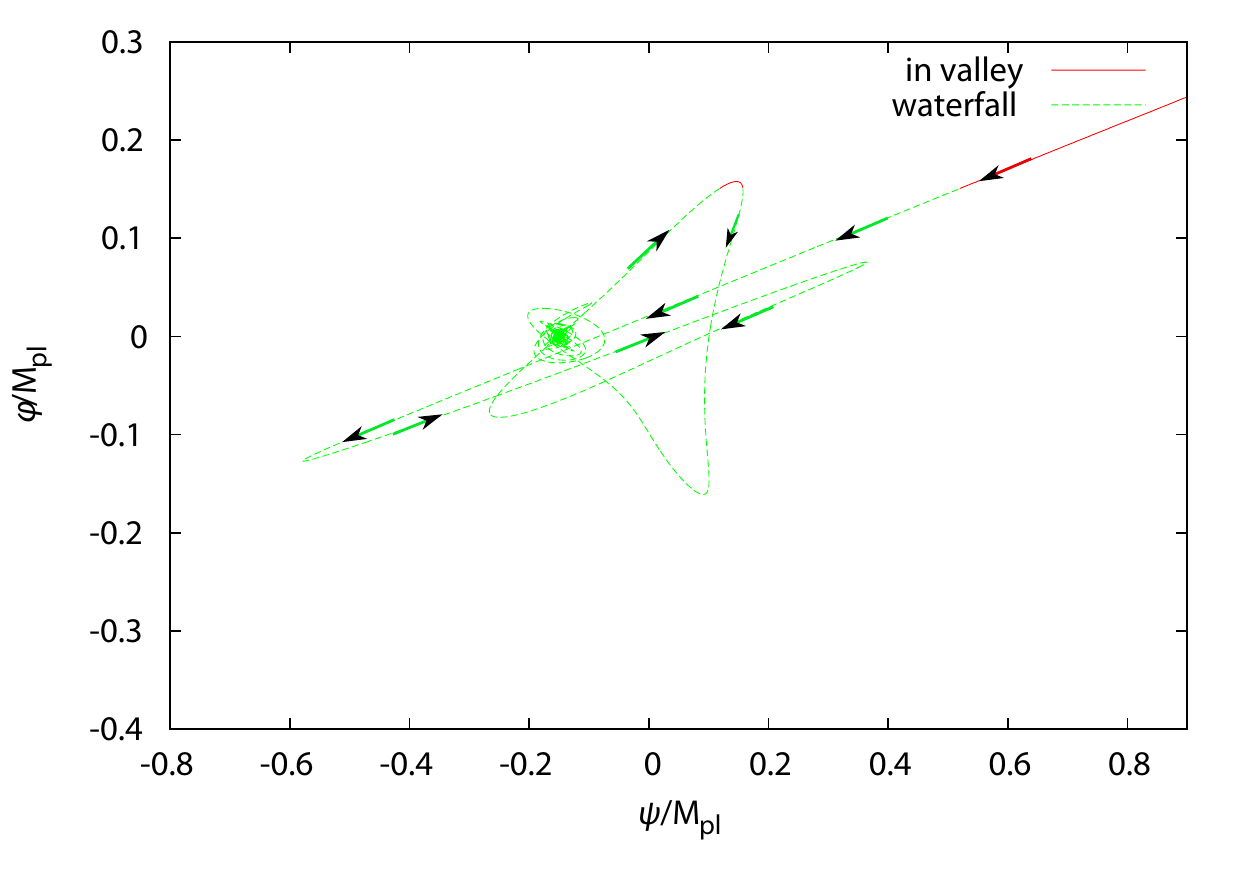}
 
  \caption{\label{fig:background-sol}Exact numerical  solution for a B-type background trajectory. (Top) Evolution of the fields with the number of e-folds realized. The inset clearly shows that this evolution is qualitatively similar to the B-type trajectory. (Bottom) Evolution of fields in configration space. It shows that the trajectory executes a complicated path and even comes out of the waterfall regime --  green in valley and red for waterfall.}
  \end{figure}
 
\subsection{Multi-field background dynamics} 

Before specializing to the hybrid case, we quickly recapitulate the formalism following Ref.~\cite{Price2015}. The Klein-Gordon equations for the homogeneous background fields $\phi_I$ where $\phi_I \, \in \{\psi, \phi, etc.\} $ take the form
\begin{equation}
\ddot{\phi}_I + 3H \dot{\phi}_I + \frac{\pd V}{\pd \phi^I} = 0.   
\label{eqn:KGtime}
\end{equation}

Replacing cosmic time $t$ with the number of $e$-folds $N_e \equiv \ln a(t)$, as it is more convenient for inflationary dynamics, gives

\begin{equation}
\frac{\ud^2 \phi_I}{\ud N_e^2 } + (3 -\epsilon) \frac{\ud \phi_I}{\ud N_e} + \frac{1}{H^2}\frac{\pd V}{\pd \phi^I} = 0  ,
\label{eqn:back}
\end{equation}
where we have defined the first slow-roll parameter as
\begin{equation}
  \epsilon\equiv-\frac{\dot H}{H^2}
  =\frac{1}{2} \frac{\ud \phi_I}{\ud N_e} \frac{\ud \phi^I}{\ud N_e}.
  \label{eqn:eps}
\end{equation}

\subsection{Perturbed equations}
We apply perturbation theory to study the spatially dependent modes, writing 
\begin{equation}
\phi_I(t, \vec{x}) \to \phi_I(t) + \delta \phi_I(t, \vec{x})
\end{equation} 
where $\phi_I(t)$ is the homogeneous, classical background field. The mode eqaution for the Fourier transformed scalar perturbation in the spatially flat gauge is
\begin{equation}
\frac{\ud^2 \delta \phi_I}{\ud N_e^2} + (3-\epsilon) \frac{\ud \delta \phi_I}{\ud N_e} + \frac{k^2}{a^2H^2} \delta \phi_I +  C_{IJ} \delta \phi^J = 0,
\label{eqn:dphi_mode}
\end{equation}
where
\begin{equation}\label{eqn:cij}
C_{IJ} \equiv \frac{\pd_I\pd_J V}{H^2} + \frac{1}{H^2} \left(\frac{\ud \phi_I}{\ud N_e} \pd_J V  +  \frac{\ud \phi_J}{\ud N_e} \pd_I V \right)
+ (3-\epsilon) \frac{\ud \phi_I}{\ud N_e} \frac{\ud \phi_J}{\ud N_e}
\end{equation}
and $\pd_I \equiv \pd/\pd \phi_I$.
Similarly the evolution equation for the tensor perturbations is written which is identical to the single-field case as tensor perturbation are massless (being non-gauge degrees of freedom) and only minimally coupled to the matter sector.

For convenience, the Mukhanov-Sasaki variable  $u_I \equiv a \delta \phi_I$ is used, and therefore the mode equation for $u_I$ becomes
\begin{equation}
\frac{\ud^2 u_I}{\ud N_e^2} + (1-\epsilon) \frac{\ud u_I}{\ud N_e} + \left(\frac{k^2}{a^2H^2} - 2 + \epsilon \right) u_I + C_{IJ} u^J = 0
\label{eqn:ptbmodeeqn}
\end{equation}
with $C_{IJ}$ as in Eq.~\eqref{eqn:cij}.
The perturbation Eqs.~\ref{eqn:ptbmodeeqn} generally mix the annihilation operators for all of the fields as the mass matrix $m^2_{IJ} \equiv \pd_I\pd_J V$ is not necessarily diagonal\cite{Salopek1989}.
Therefore each perturbation mode $u_I(\bk)$ and $u_I^\dagger(\bk)$ is expanded using $N$ harmonic oscillators $a_J(\bk)$:
\begin{equation}
  u_I(\bk, N_e) = \psi_{I}^{\;\; J} (\bk, N_e) a_J(\bk) \qquad \mathrm{and} \qquad
u_I^\dagger(\bk, N_e) = \psi_{I}^{\;\;J, *}(\bk, N_e) a_J^\dagger(\bk) ,
\end{equation}
where $N$ is the number of fields in the potential and $(\dagger)$ and $(*)$ represent Hermitian and complex conjugation, respectively.
Then one can write canonical commutation relations $ [ a_J(\bk), a^\dagger_I(\bk')] = (2\pi)^3 \delta_{IJ} \delta^{(3)}(\bk - \bk')$. Finally the mode matrix $\psi_{IJ}$ evolves according to
\begin{equation}
\frac{\ud^2 \psi_{IJ}}{\ud N_e^2} + (1-\epsilon) \frac{\ud \psi_{IJ}}{\ud N_e} + \left(\frac{k^2}{a^2H^2} - 2 + \epsilon \right) \psi_{IJ} + C_{IL} \psi^{L}_{\; \, J} = 0 .
\label{eqn:psi}
\end{equation}
We set the Bunch-Davies initial conditions~\cite{Bunch1978} in Eq.~\eqref{eqn:psi} to get the perturbation spectrum. First, the time at which the pivot scale $k_*$ leaves the horizon, $N_*$ $e$-folds before the end of inflation, is found using the background equations~\eqref{eqn:back}. Then the times  $N_{e,\mathbf{k}}$ when other $\mathbf k$ modes leave the horizon are calculated.

The Bunch-Davies state~\cite{Salopek1989a} chooses field bases such that the $\psi_{IJ}$ are originally diagonal and sets the initial condition for Eq.~\eqref{eqn:psi} as if the mode matrix were freely oscillating in Minkowski space. It is to be noted that for a single field, mode evolution is ended when the mode is frozen out of the horizon. On the other hand, to include superhorizon evolution, modes for multifield models are evolved until the end of inflation~\cite{Price2015}.

\subsection{Power spectra}~\label{subsec:powerspectra}

In multifield models, contractions of the mass matrix  are required to obtain the power spectrum. 
Using the canonical commutation relations above, the two-point VEV of the field perturbations yields the power spectrum
\begin{equation}
  P_{\delta \phi}^{IJ} (k) = \frac{k^3}{2\pi^2} \left[\frac{1}{a^2}\right] \psi^{I}_{\; L} \; \psi^{J L,*} \, .
  \label{eqn:power}
\end{equation}

To relate this field-space power spectrum to gauge-invariant perturbation variables~\cite{Bardeen1980,Bardeen1983,Nibbelink2002}, the curvature perturbation $\calR$  is defined on comoving hypersurfaces by
\begin{equation}
  \calR \equiv \Psi + \frac{1}{3} \nabla^2 E + aH \left( B + v \right) ,
  \label{eqn:rginvariant}
\end{equation}
where $E$ and $B$ are perturbations to the flat FLRW metric, and
$v$ is given in terms of the momentum density of the stress-energy tensor $T^{\mu}_{\; \; \nu}$ as
\begin{equation}
  T^{i}_{\; \, 0} \equiv \left( \bar \rho + \bar P \right) \delta^{ij} \frac{\pd v}{\pd x^j}  ,
  \label{eqn:XXX}
\end{equation}
where $\bar \rho$ and $\bar P$ are the background energy and pressure densities, respectively.
On evaluating Eq.~\eqref{eqn:rginvariant} on spatially-flat hypersurfaces during inflation, $\calR$ reduces to
\begin{equation}
  \calR = -\frac{H}{\dot \phi_0} \; \omega_I \delta \phi^I  ,
  \label{eqn:rptb}
\end{equation}
where $\omega_I \equiv \dot \phi_I/ \dot \phi_0$ is a basis vector that projects $\delta \phi_I$ along the direction of the classical background trajectory, given by the solutions to Eq.~\eqref{eqn:back}.  The vector $\vec \omega$ and a complementary set of $(N_f-1)$ mutually orthonormal basis vectors $\vec s_K$ form the kinematic basis, where the separation between the adiabatic perturbations in Eq.~\eqref{eqn:rptb} and transverse, isocurvature perturbations is made explicit.  Since $\vec \omega$ depends on the nonlinear background evolution, we find the $\vec s_K$ numerically by Gram--Schmidt orthogonalization.

The \emph{adiabatic curvature power spectrum} $\PR$ is then the projection of $\PIJ$ along the field vector $\omega_I$, scaled by the pre-factor in Eq.~\eqref{eqn:rptb}, leaving
\begin{equation}
  \PR (k) = \frac{1}{2 \epsilon} \omega_I  \omega_J \PIJ(k)  .
  \label{eqn:pad}
\end{equation}
The gauge-invariant scalar density spectrum in Eq.~\eqref{eqn:pad} is the final result for the adiabatic two-point function to first-order in perturbation theory.

To determine the scalar power spectrum at the end of inflation in hybrid inflation, it is common usage to restrict the dynamics to
the effective one-field potential along the valley $\psi=0$
\begin{equation} \label{eq:potenhybeffectif}
V_{\text eff}(\phi) =  \Lambda^4 \left[1+ \left( \frac{\phi}{\mu}
\right)^2\right],
\end{equation}
and to assume that inflation ends abruptly once the instability
point is reached. Under these hypotheses, the primordial
scalar power spectrum can be easily derived in the
slow-roll approximiation. It is nearly scale invariant,
with a spectral tilt
\begin{equation}
n_\mathrm{s} -1 \equiv \left.\frac{\dd\PR}{\dd \ln k}\right|_{k=k_*} = - 2 \epsilon_{1*} -\epsilon_{2*}~.
\end{equation}
where a star means that the quantity is evaluated when
the pivot mode $k_*$ 
leaves the Hubble radius, that is when  $k_*=aH$.


\section{Numerical solutions}~\label{sec:numerics}

We stop integrating when  either (a) the orbit achieves more than 60 e-folds during inflation or (b) $\rho < \Lambda^4$ and the trajectory is trapped by the potential wells at $\{\psi,\phi\} = \{\pm M, 0 \}$. 

Although there exist other types of trajectories such A, B, C and D as classified in \cite{Clesse2009}, we have focussed only on the B-type trajectory.

 \begin{figure} 
 \resizebox{0.5\textwidth}{0.38\textwidth}{\includegraphics{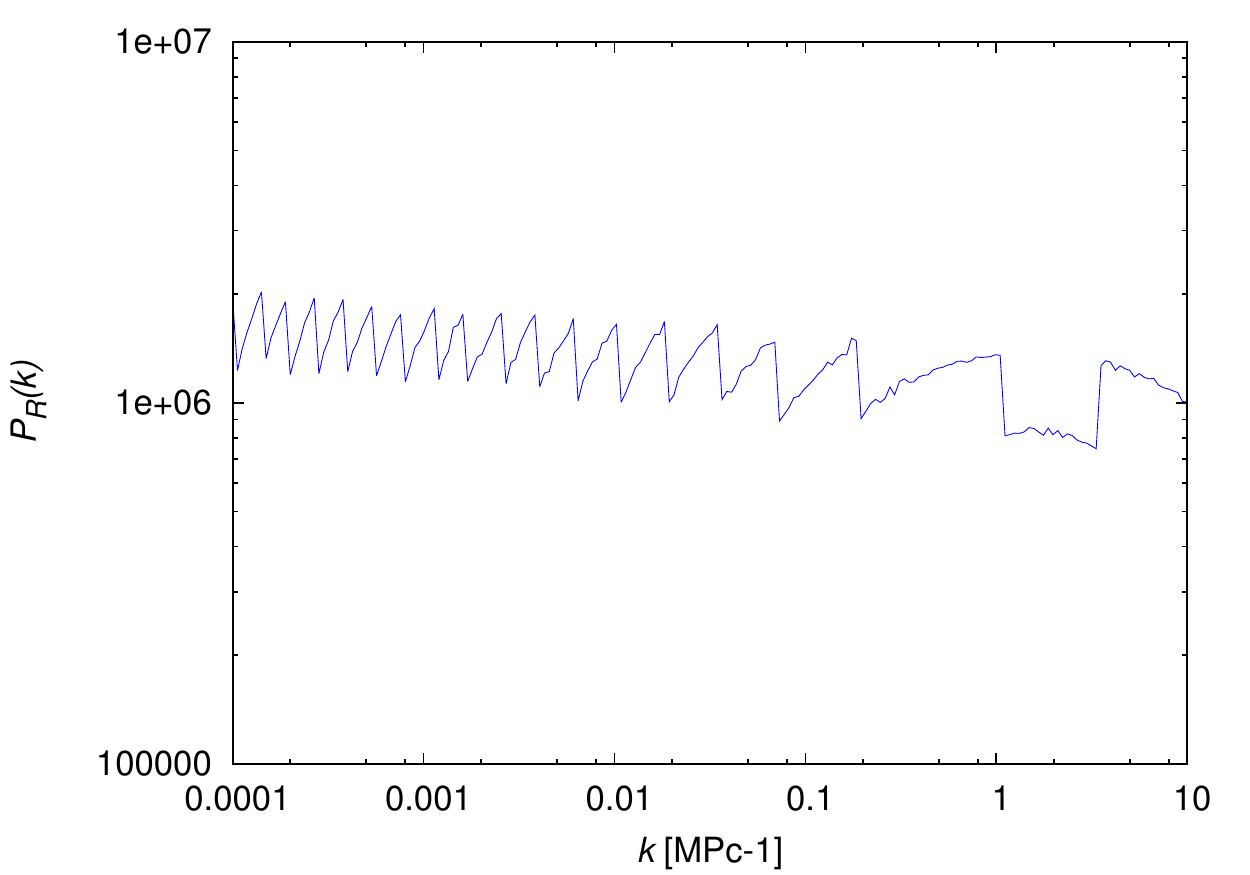}}
 \resizebox{0.5\textwidth}{0.4\textwidth}{\includegraphics{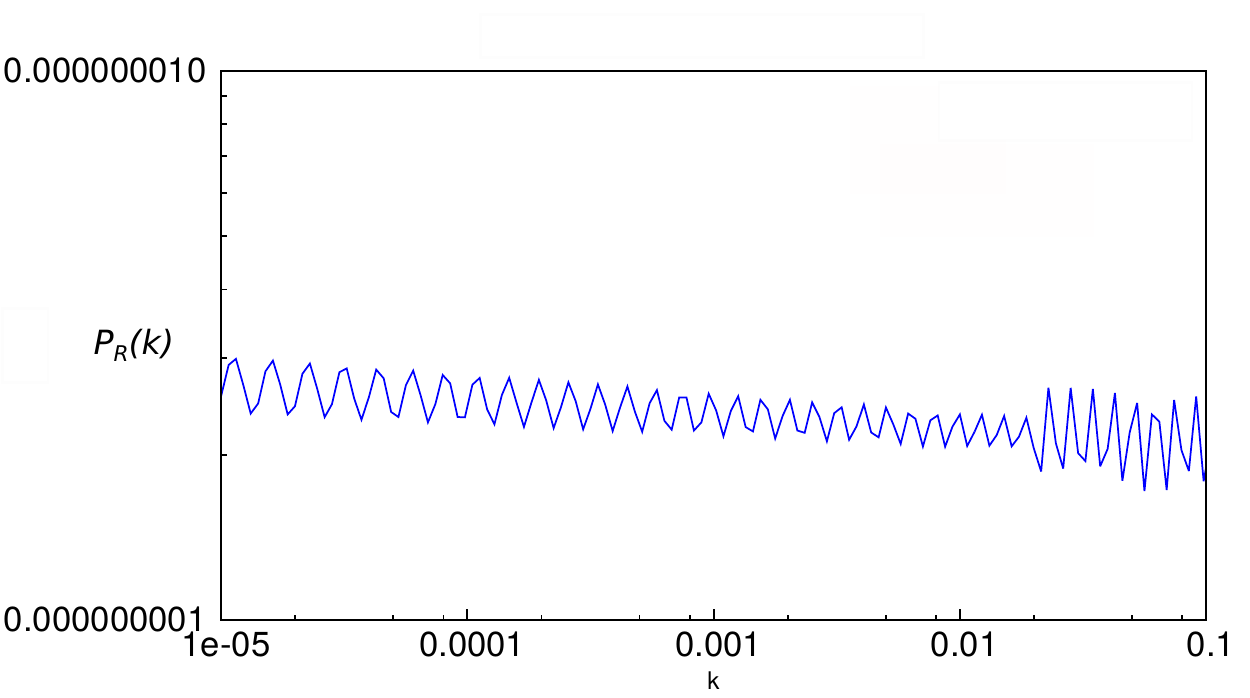}}
 \caption{\label{fig:prk-axion-monodromy} Oscillatory features in the power spectra. The spectrum of the hybrid potential (left) contains the drifting oscillatory features which are similar to that in the spectrum of axion  model (right).}
 \end{figure}


Setting $\Lambda = 0.1064 \, \mpl$, $M=0.1504  \, \mpl$, $\mu=3190.4  \, \mpl$, and $\nu=0.1265 \, \mpl$ in 
 \begin{equation}
   \label{eqn:powerspect}
   A_s \approx \frac{1}{24\pi^2 M_\mathrm{Pl}^4} \left( \frac{V}{\epsilon_V} \right),
 \end{equation}
 where $\epsilon_V = (M_\mathrm{Pl}^2/2)(V_{,\phi}/V)^2$ is the slow-roll parameter, and assuming perturbations are generated when $\psi \approx 0$ and $\phi \approx \phi_c$,  $ A_s$ turns out to be $ \approx 1.2\e{9} \, \mpl$.\footnote{Generally the amplitude $A_s$ of the dimensionless power spectrum $\mathcal{P}_\mathcal{R}$ is set to be roughly compatible with the WMAP9 data \cite{Benetti2011,Hinshaw2013}, which fixes the potential energy scale.}  Ref.~\cite{Martin2012} determined that quantum fluctuations dominate the classical field evolution in the inflationary valley when
 \begin{equation}
   \label{eqn:lambdaquant}
   \Lambda > \Lambda_q \text{ where } \Lambda_q^2 \equiv 4 \pi \sqrt{3} M_\mathrm{Pl}^3 \frac{\phi_c}{\mu^2}.
 \end{equation}
 Although, for our parameters $\Lambda_q \simeq 5.7\e{-4} \, \mpl$ \textit{i.e.} $\Lambda > \Lambda_q$, perturbations are generated outside the valley \textit{i.e.} when $ |\psi| \simeq 0.075 $. Therefore for the purpose of perturbation spectrum, we will assume that classical equations are self-consistent.
 
 \subsection{Oscillatory features}

Searching for oscillations is well-motivated, because a discovery would
 imply a very interesting additional structure in the primordial perturbations, and conversely
 a null result can constrain the parameter space in a useful way~\cite{Flauger2014}. 
As is clear from Figure~\ref{fig:prk-axion-monodromy}, oscillatory features in $\PR(k)$ for the hybrid inflation potential are qualitatively similar to those of the axion monodromy potential.

 \begin{figure}[!t]
  \resizebox{0.5\textwidth}{!}{\includegraphics{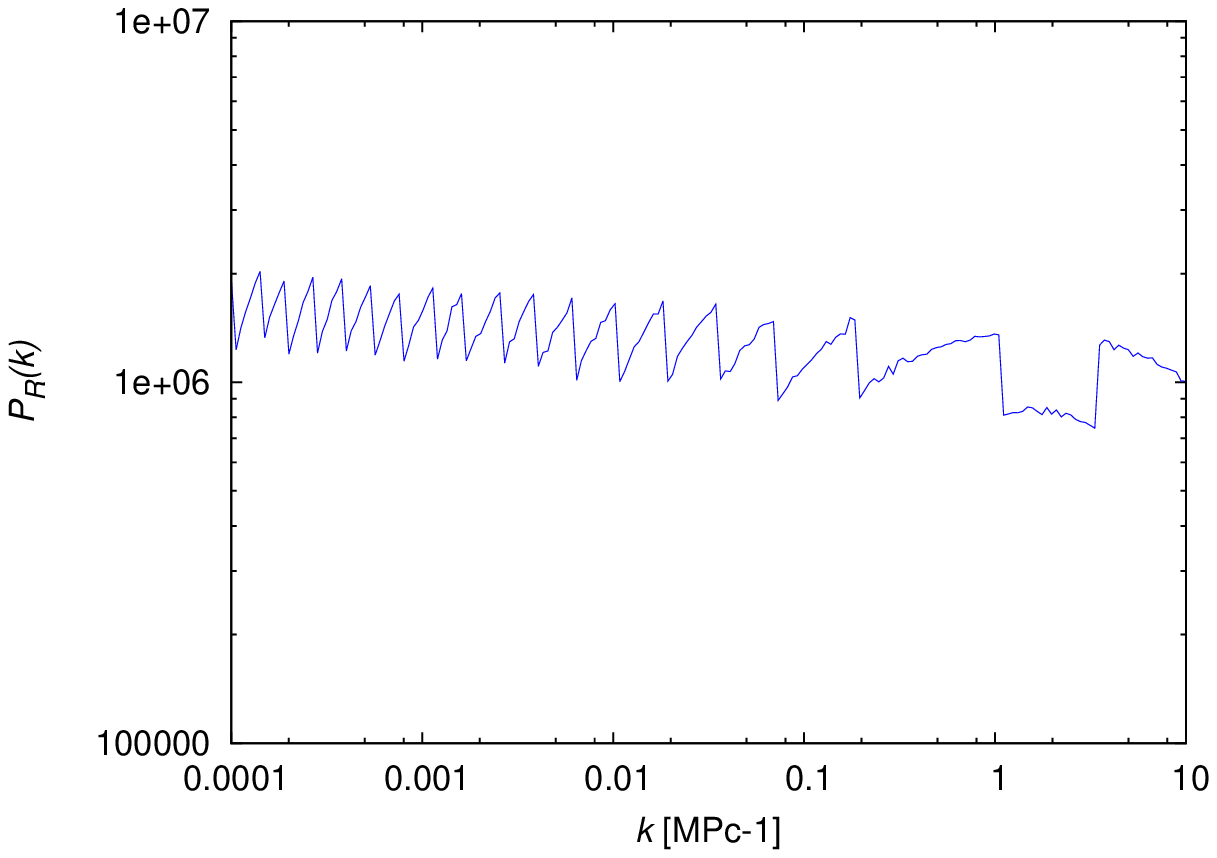}}
  \resizebox{0.5\textwidth}{!}{ \includegraphics{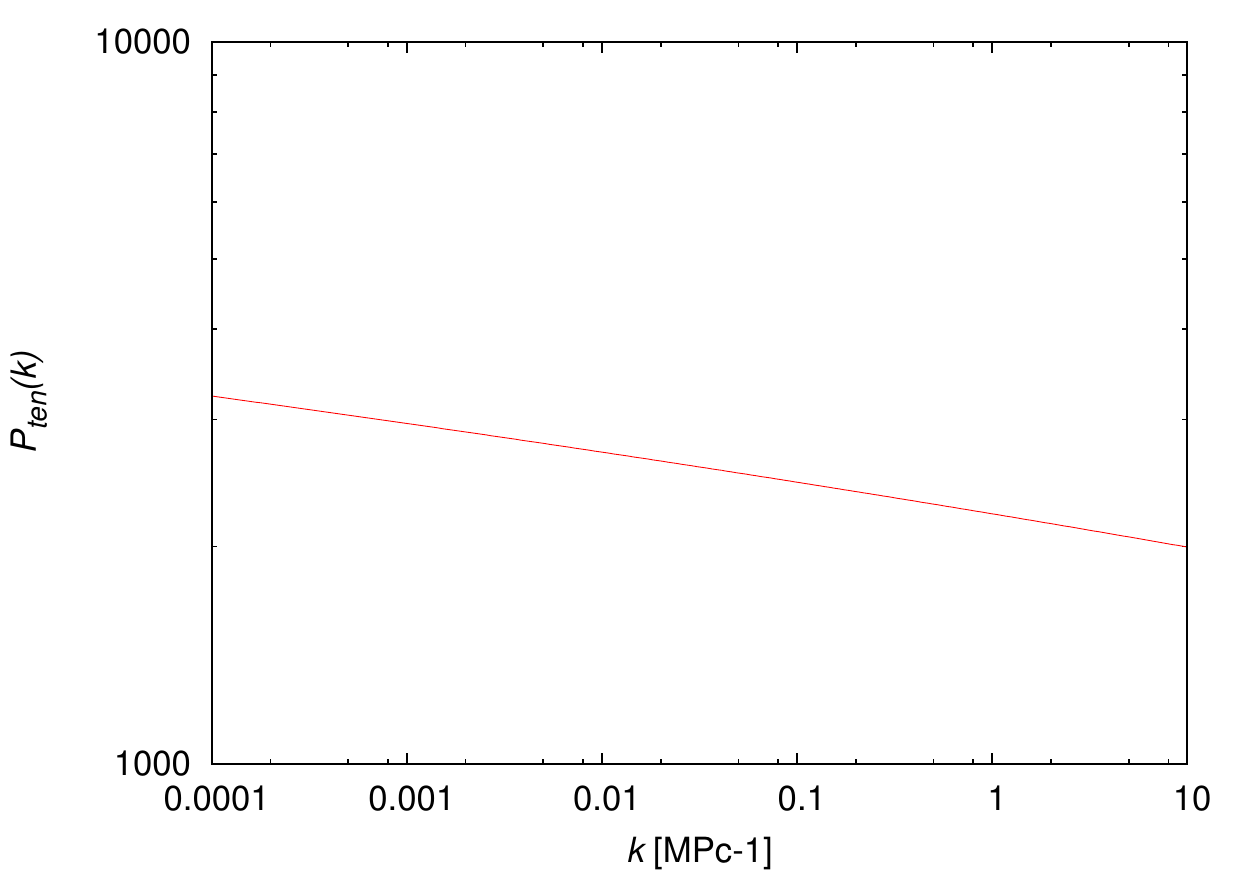}}
  \caption{\label{fig:k-dependent r}$\PR(k)$ versus $\Pten(k)$ in hybrid model. The tensor spectrum varies very little, whereas adiabatic $\PR(k)$ contains oscillations, and thus rendering $r$ strongly $k$-dependent.}
  \end{figure}
   
 The inflaton potential in the axion monodromy model  motivated by power-law moduli stabilization  is given by
 \begin{equation}
 V(\phi)
 =\lambda\, \left[\phi+\alpha\,\cos\left(\frac{\phi}{\beta}+\delta\right)\right].
 \label{eqn:amm}
 \end{equation} 
 Evidently, in this potential, while the parameters~$\alpha$ and
 $\beta$ describe the amplitude and the frequency of the superimposed 
 oscillations, the parameter~$\delta$ shifts the oscillations within 
 one period. Further the amplitude of the oscillation is fixed in the axion monodromy model and 
 the inflaton oscillates as it rolls down this potential, and these
 oscillations persist all the way until the end of inflation~\cite{Hazra2013}. 
 Oscillatory contributions to the scalar power spectrum  arise as a direct consequence of the underlying periodicity. However, the oscillatory features  have a model-dependent amplitude, and may be undetectably small: in particular, the amplitude is exponentially suppressed in regimes where the oscillations are generated by instanton
 effects~\cite{Flauger2010, Flauger2014}.
 
For the inflaton potential (\ref{eqn:amm}), using the same approximations as in \cite{Flauger2010}, the power spectrum can be fitted by a function of the form
 \begin{equation}\label{eq:Ppf}
 \Delta_\mathcal{R}^2(k)=\Delta_\mathcal{R}^2\left(\frac{k}{k_\star}\right)^{n_s-1}\left(1+\delta n_s e^{-C_0\left(\frac{\phi_k}{\phi_0}\right)^{p_\Lambda}}\cos\left[\frac{\phi_0}{f}\left(\frac{\phi_k}{\phi_0}\right)^{p_f+1}+\Delta\varphi\right]\right)\,,
 \end{equation}
 where the amplitude $\delta n_s$ and the phase $\Delta\varphi$ are calculable using the techniques in~\cite{Flauger2010}; $\phi_0$ is some fiducial value for the scalar field to be specified later; and $\phi_k$ represents the scalar field value at the time when the mode with comoving momentum $k$ exits the horizon~\cite{Flauger2014}. Along similar lines, a fitting function for the spectrum of the hybrid potential needs to be developed, which we will undertake in future work.
In Figure~\ref{fig:k-dependent r}, we observe that $\PR (k)$  is highly oscillatory  whereas $\Pten (k)$  is featureless. It implies that tensor-to-scalar ratio $r$ for this model is strongly $k$-dependent.

   \begin{figure}[!t]
   \begin{center}
    \includegraphics{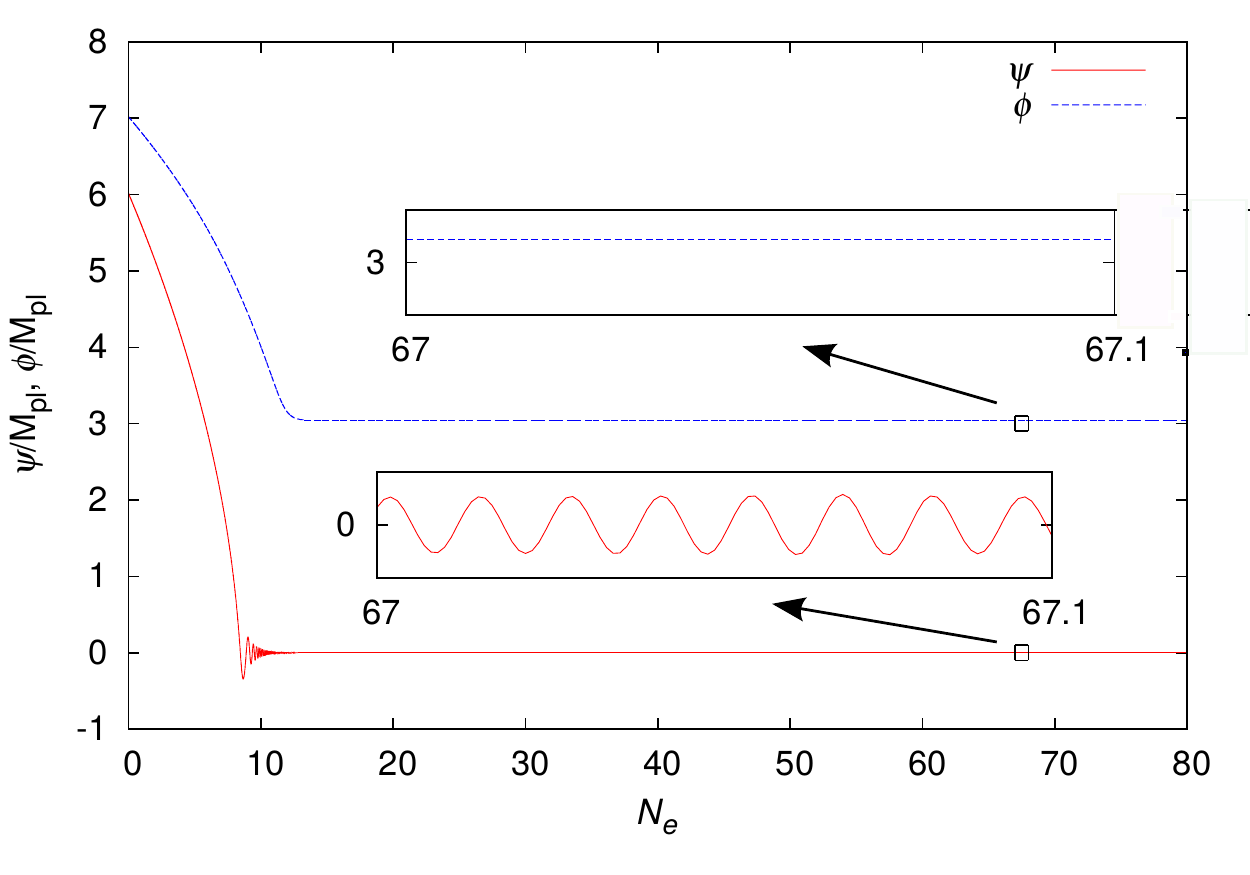} 
   
   \caption{\label{fig:A-type}Exact numerical  solution for A-type background trajectory. Respecive zooms in inset clearly show that $\psi$ is nearly stable at $\phi=\phi_{\text valley}\approx3.0464$, whereas $\psi$ oscillates in the narrow band of $\Delta \psi\approx10^14$. }
   \end{center}
   \end{figure}
    
\subsection{Numerical issues}

 There exists a numerical issue related to calcualting the power spectrum for the A-type trajectory~\cite{Clesse2009}. It is particularly tricky to numerically calculate the spectrum in this case as the oscillations along the $\psi$ direction continue in the narrow band of $\Delta \psi$, whereas  $\psi$ is nearly stable at $\phi=\phi_{\text valley}$ (3.0464 for our choice of parameters). See the zoom of A-type trajectory in Figure~\ref{fig:A-type}.
 
  \begin{figure*}[!t]
  \begin{center} 
  \resizebox{0.45\textwidth}{!}{\includegraphics{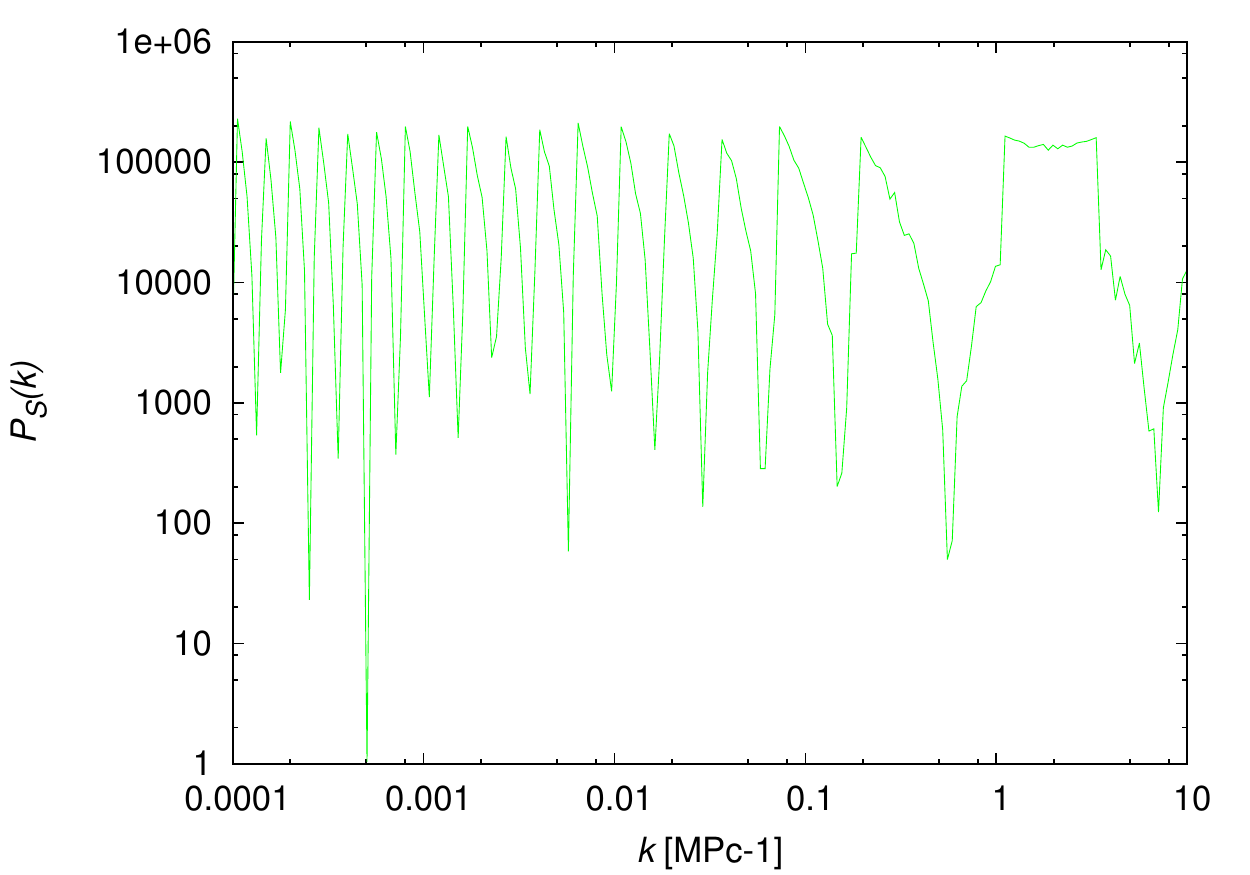}}
  \resizebox{0.45\textwidth}{!}{\includegraphics{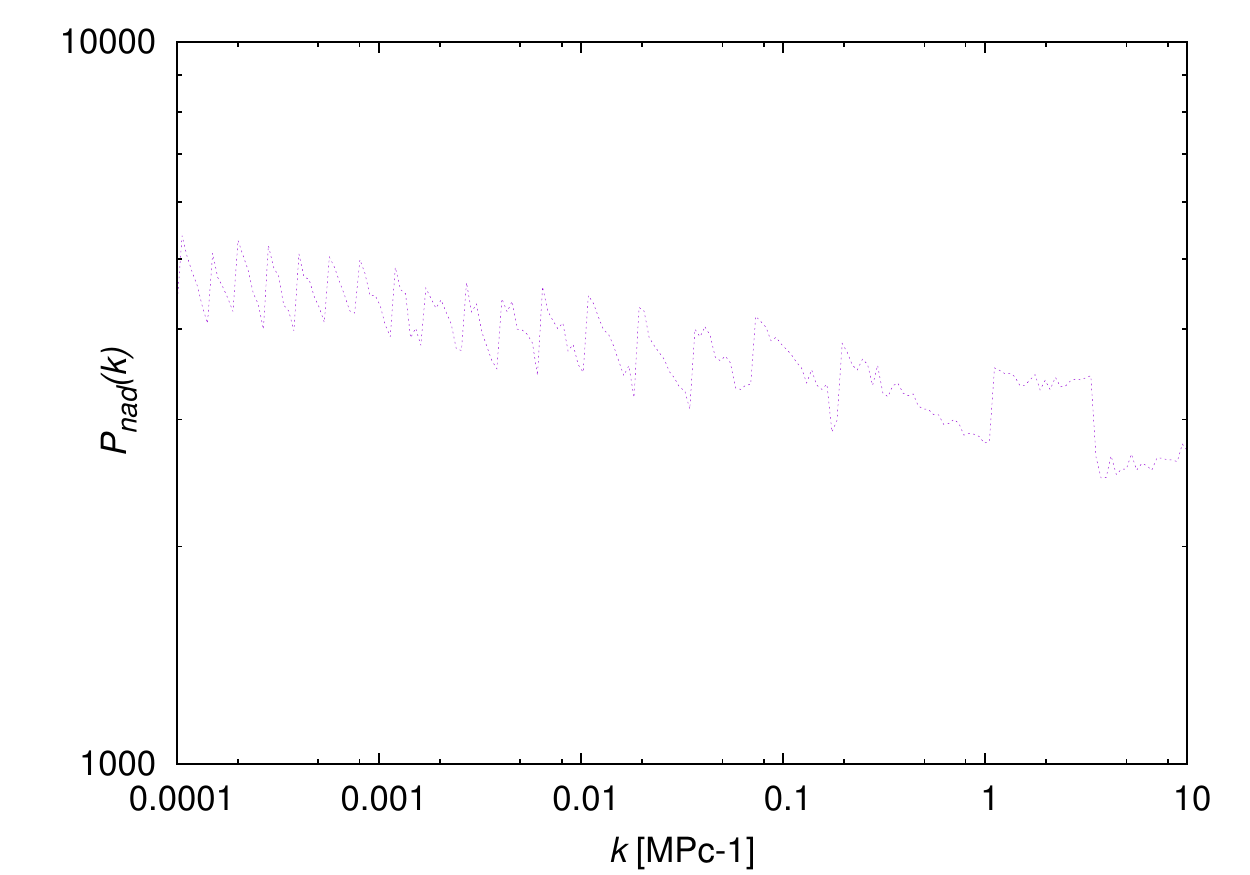}} 
  \end{center}
  \caption{\footnotesize\label{fig:entropic_mode}  Entropic $\PS(k)$ and non-adiabatic pressure $\Ppnad$ spectra. Scaled $\Ppnad$ is similar to  $\PR$.}
  \end{figure*}

\section{Conclusion and discussion}~\label{sec:conclusion}
For the original hybrid inflation model and some of its extensions, Clesse~\cite{Clesse2014} found time evolution of $\PR$ for only the pivot scale $k_*$. On the other hand, we have calculated the complete spectrum $\PR (k)$ for the original hybrid model.
Unlike in the axion model there is no underlying periodicity in the hybrid potential, but the interaction term produces the same kind of oscillatory behaviour in inflaton evolution. Therefore some characterstic features in spectrum of the hybrid potential need to be looked for to distinguish between the axion and hybrid models. It would be interesting to observe how CMB anisotropy spectra will behave for this oscillatory power spectrum. Further it can be explored how the perturbation spectrum will appear if quantum fluctuations are taken into account. Also there exists a need to develop some other theoretical techniques to calculate $\PR (k)$ for other initial conditions where numerical techniques are failing, such as in the case of A-type trajectory. We plan to examine these questions in future work.

\acknowledgments
 
One of the authors (RK) thanks Layn C. Price, Richard Easther, Sebastian Clesse and Dhiraj Kumar Hazra for many helpful communications.

\bibliographystyle{JHEP}

\bibliography{hybrid4}
 
\end{document}